\documentclass[12pt]{article}
\usepackage{epsfig,amssymb,amsfonts}
\title{\bf Nonperturbative dynamics in the colour--magnetic QCD vacuum}
\author{A. V. Nefediev and  Yu. A. Simonov}
\date{\small\it Institute of Theoretical and Experimental Physics,\\
117218, B.Cheremushkinskaya 25, Moscow, Russia}
\newcommand{\be}{\begin{equation}}
\newcommand{\ee}{\end{equation}}

\begin{document}
\maketitle
\begin{abstract}
In the deconfinement phase of
QCD quarks and gluons interact with the dense stochastic colour--magnetic
vacuum. We consider the dynamics of quarks in this deconfinement phase
using the Field Correlators Method and derive an effective nonperturbative
inter-quark potential, in addition to the usual perturbative short--ranged interaction. We find the
resulting angular--momentum--dependent interaction to be attractive enough to maintain bound states and, for light quarks
(and gluons), to cause emission of quark and gluon pairs.
Possible consequences for the strong interacting quark--gluon plasma are briefly discussed.
\end{abstract}

\maketitle

\section{Introduction}

The picture of the strong interacting quark--gluon plasma (SQGP) seems to be adequate for explaining the recent data on ion--ion collisions
\cite{1}. It was found on the lattice \cite{2} that colour--magnetic vacuum fields do not change across the phase transition, thus
supporting the conjecture made in Ref.~\cite{3} that, above the temperature of deconfinement $T_c$, the QCD vacuum loses its confining
colour--electric part, while the colour--magnetic part remains intact. This idea can be economically expressed in the formalism of
the Field Correlators Method (FCM) \cite{4}, where the Gaussian correlators of the colour--electric and colour--magnetic fields,
$\langle\langle E_i(x) E_j(y)\rangle\rangle$ and $\langle\langle H_i(x) H_j(y)\rangle\rangle$
($\langle\langle\ldots\rangle\rangle$ denotes irreducible correlators), are parametrised through the correlation
functions $D^E$, $D^E_1$, $D^H$, and $D^H_1$, respectively (see Ref.~\cite{4} for the details of the formalism)\footnote{The correlators $D^E$
and $D^H$ appear at the structure $\delta_{ij}$ in these colour--electric and colour--magnetic correlators, respectively
(see also Eq.~(\ref{FF})), whereas $D_1^{E,H}$ parametrise the terms with derivatives.}. Notice that
among those only  $D^E$ vanishes above the $T_c$ \cite{2,3}.
The usual (time-like) string tension $\sigma_E$ and the so-called spatial string tension $\sigma_s$ are given by the double integrals from the
corresponding correlators,
\be
\sigma_E=\frac12 \int D^E (\xi) d^2\xi,\quad\sigma_s\equiv\sigma_H=\frac12\int D^H(\xi)d^2 \xi.
\label{sigmas}
\ee
Both string tensions coincide below the $T_c$ but
$\sigma_E$ is expected to vanish in the deconfinement phase, whereas $\sigma_H$ remains nearly constant in the vicinity of the critical
temperature, both below and above the $T_c$ \cite{5,6}.
Moreover, $\sigma_H$ grows at large $T$, as $\sqrt{\sigma_H}\propto Tg^2 (T)$ \cite{6}, signalising that $D^H$ grows as ${\cal O}(T^2
g^4(T))$. At the same time it was conjectured in Ref.~\cite{7}, and confirmed later on the lattice \cite{2,8}, that the
non-confining correlator $D_1^E$ does not vanish either above the $T_c$ and leads to strong interaction between quarks and gluons.
In Refs.~\cite{8,9}, bound states due to $D^E_1$ in quark and gluonic systems where found in qualitative agreement with the lattice data,
Ref.~\cite{10}. Many efforts based on the colour--electric type forces have been applied to clarify the dynamics of QCD above the
deconfinement phase transition. Still, without the colour--magnetic forces, the dynamical picture of the
SQGP is not complete
since colour--electric forces cannot bind quarks and gluons for large angular momentum
where colour--magnetic forces are most important. Thus the purpose of the present Letter is to clarify
the matter and to study orbitally excited bound states of quarks above the $T_c$.
For the sake of simplicity, we consider only the
confining correlators $D^{E,H}$ and neglect the other two, $D_1^{E,H}$.
Furthermore, we study a strongly interacting system at the fixed temperature $T$, which can be either below or above the deconfinement
temperature $T_c$, and develop the standard Hamiltonian approach to it.
We consider the situation when the colour--electric interaction is switched off above the deconfinement temperature and the dynamics of quarks and
gluons is governed by the colour--magnetic forces only. Finally, we
use the interaction under consideration to estimate the strength of the interaction in
the SQGP by comparing the mean potential energy of light quarks (and
gluons) in the plasma with their mean kinetic energy. We find the corresponding parameter to be of order $\sigma_H/T^2$. Since this parameter
is large for quarks and even $9/2$ times larger for gluons we conclude
that SQGP is very strongly coupled and it should be viewed
as a liquid, at least.

\section{QCD string and the spin--independent interaction in the quark--antiquark system}

In this section we derive the effective spin--independent quark--antiquark interaction arising due the
QCD string formation.

Following the method proposed in Refs.~\cite{11,DKS} we consider the gauge--invariant Green's function in the vacuum at temperature $T$:
\be
G(x_1,x_2|y_1,y_2)=\langle\Psi_{out}(x_1,x_2)\Psi_{in}^\dagger(y_1,y_2)\rangle,
\label{Gf}
\ee
where the wave functions of the initial and final colourless $q\bar{q}$ states are built with the help of the parallel transporter
$\Phi(x,y)=P\exp\left[i\int_y^xdz_{\mu}A_{\mu}(z)\right]$. Using the standard path integral approach and the Feynman--Schwinger
representation for the single--quark propagator, one arrives at the Green's function (\ref{Gf}) in the
form \cite{11}:
\be
G(x_1,x_2|y_1,y_2)=\int_0^\infty ds \int (Dz)^w_{x_1y_1}e^{-K}\int_0^\infty d\bar{s}\int (D\bar{z})^w_{x_2y_2}e^{-\bar{K}}
\langle TrW(C)\rangle,
\label{1}
\ee
with the kinetic energies $K=\frac14\int^s_0 d\tau\dot{z}_\mu^2$ and
$\bar{K}=\frac14\int^{\bar{s}}_0 d\tau\dot{\bar{z}}_\mu^2$
and with the ``winding path measure'' $(Dz)^w_{xy}$ taking into account Matsubara periodic boundary conditions.
All spin--dependent terms are neglected in Eq.~(\ref{1}) --- they will be restored below, in
Section~\ref{SDint}. The closed contour $C$ runs over the
quark trajectories and the dynamics of the system is defined by the averaged Wilson loop $\langle TrW(C)\rangle$.
In the Gaussian approximation, one finds that \cite{4}
\be
\frac{1}{N_C}\langle Tr W(C)\rangle=\exp\left(-\frac12\int_Sd\sigma_{\mu\nu}(x)\int_Sd\sigma_{\lambda\rho}(x')
\langle\langle Tr F_{\mu\nu}(x)\Phi(x,x')F_{\lambda\rho}(x')\Phi(x',x)\rangle\rangle\right),
\label{W}
\ee
where $S$ is the minimal surface bounded by the contour $C$.
Keeping only the string--generating field strength correlators, we have
\be
\langle\langle Tr F_{\mu\nu}(x)\Phi(x,x')F_{\lambda\rho}(x')\Phi(x',x)\rangle\rangle=
(\delta_{\mu\lambda}\delta_{\nu\rho}-\delta_{\mu\rho}\delta_{\nu\lambda})D((x-x')^2).
\label{FF}
\ee

In what follows we distinguish between the electric and magnetic contributions in Eq.~(\ref{FF}), so that the structure
$d\sigma_{0i}d\sigma_{0i}$ enters Eq.~(\ref{W}) multiplied by $D^E$, whereas the spatial part $d\sigma_{jk}d\sigma_{jk}$ is accompanied by
$D^H$. The electric and magnetic string tensions are defined then according to Eq.~(\ref{sigmas}).

We synchronise the quarks in the laboratory frame, putting $x_{10}=x_{20}=t$, and adopt the straight--line ansatz for the minimal string writing
\be
d\sigma_{\mu\nu}(x)=\varepsilon^{ab}\partial_a w_\mu(t,\beta)\partial_b w_\nu(t,\beta)dt d\beta,\quad \{a,b\}=\{t,\beta\},
\ee
where $0\leqslant t\leqslant {t_{\rm max}}$, $0\leqslant\beta\leqslant 1$, and the profile function is defined by the trajectories of the quarks,
$w_\mu(t,\beta)=\beta x_{1\mu}+(1-\beta)x_{2\mu}$. For further convenience let us introduce two vectors:
\be
\vec{r}=\vec{x}_1-\vec{x}_2,\quad\vec{\rho}=[(\vec{x}_1-\vec{x}_2)\times(\beta\dot{\vec{x}}_1+(1-\beta)\dot{\vec{x}}_2)]\equiv r\vec{\omega},
\ee
which allow one to write the differentials in a compact form,
\be
d\sigma_{0i}(x)d\sigma_{0i}(x')=\vec{r}(t)\vec{r}(t')dtdt'd\beta d\beta',\quad
d\sigma_{jk}(x)d\sigma_{jk}(x')=2\vec{\rho}(t)\vec{\rho}(t')dtdt'd\beta d\beta'.
\ee

Presenting the averaged Wilson loop as
$$
\frac{1}{N_C}\langle Tr W(C)\rangle=e^{-J},\quad J=J^E+J^H,
$$
one can write for the electric and magnetic contributions separately:
\begin{eqnarray}
J^E=\int_0^{t_{\rm max}}dt\;dt'\int_0^1d\beta\;d\beta'\;\vec{r}(t)\vec{r}(t')D^E((x-x')^2),\label{JE}\\
J^H=\int_0^{t_{\rm max}}dt\;dt'\int_0^1d\beta\;d\beta'\;\vec{\rho}(t,\beta)\vec{\rho}(t',\beta')D^H((x-x')^2)\label{JB}.
\end{eqnarray}

The correlation functions $D^{E,H}$ decrease in all directions of the Euclidean space--time with the correlation length $T_g$ which is
measured on the lattice to be rather small, $T_g\approx 0.2\div 0.3$~fm \cite{camp}. Therefore, only close points $x$ and $x'$ are correlated, so that one can neglect the difference
between $\vec{r}(t)$ and $\vec{r}(t')$, $\vec{\rho}(t,\beta)$ and $\vec{\rho}(t',\beta')$ in Eqs.~(\ref{JE}), (\ref{JB}) and also write:
\be
(x-x')^2=(x(t,\beta)-x(t',\beta'))^2=g^{ab}\xi_a\xi_b,\quad\xi_a=t-t',\quad\xi_b=\beta-\beta'.
\ee
The induced metric tensor is $g^{ab}=g^a\delta^{ab}$, $g^1g^2=det\;g=r^2+\rho^2=r^2(1+\omega^2)$. Now, after an appropriate change of variables and
introducing the string tensions, according to Eq.~(\ref{sigmas}), one readily finds:
\begin{eqnarray}
J^E=\sigma_E \int_0^{t_{\rm max}}dt\int_0^1d\beta\frac{r^2}{\sqrt{r^2+\rho^2}}=\sigma_E r\int_0^{t_{\rm max}}dt\int_0^1d\beta\frac{1}{\sqrt{1+\omega^2}},\\
J^H=\sigma_H \int_0^{t_{\rm max}}dt\int_0^1d\beta\frac{\rho^2}{\sqrt{r^2+\rho^2}}=\sigma_H r\int_0^{t_{\rm max}}dt\int_0^1d\beta\frac{\omega^2}{\sqrt{1+\omega^2}}.
\end{eqnarray}

For $\sigma_E=\sigma_H=\sigma$ the sum of $J^E$ and $J^H$ reproduces the well-known action of the Nambu--Goto string.

For further analysis we resort, as was done in Ref.~\cite{DKS}, to the Hamiltonian description of the quark--antiquark system under
consideration. We also turn over to Minkowski space--time.

Then the Lagrangian of the quark--antiquark system can be derived from the exponent in Eq.~(\ref{1}) in the form:
$$
L=-m_1\sqrt{1-\dot{\vec{x}}_1^2}-m_2\sqrt{1-\dot{\vec{x}}_2^2}
-\sigma_Er\int_0^1d\beta\frac{1}{\sqrt{1-[\vec{n}\times(\beta\dot{\vec{x}}_1+(1-\beta)\dot{\vec{x}}_2)]^2}}
$$
\be
+\sigma_Hr\int_0^1d\beta\frac{[\vec{n}\times(\beta\dot{\vec{x}}_1+(1-\beta)\dot{\vec{x}}_2)]^2}
{\sqrt{1-[\vec{n}\times(\beta\dot{\vec{x}}_1+(1-\beta)\dot{\vec{x}}_2)]^2}},\quad \vec{n}=\frac{\vec{r}}{r}.
\label{L1}
\ee
Two particular cases of the Lagrangian (\ref{L1}) are of most interest. The first such case corresponds to equal masses, whereas in the other
case one mass is assumed infinitely large. Then, using the standard technique,
one can proceed to the Hamiltonian of the quark--antiquark system,
\be
H=\frac{1}{\xi}\left(\frac{p_r^2+m^2}{\mu}+\mu\right)+\int^1_0d\beta\left(\frac{\sigma_1^2r^2}{2\nu}+
\frac{\nu}{2}+\sigma_2 r\right)+\frac{\vec{L}^2}{r^2[\xi\mu+2\int^1_0d\beta\nu(\beta-\xi/2)^2]},
\label{Hgen}
\ee
where $\xi=1$ for the case of equal masses ($m_1=m_2=m$) and $\xi=2$ for the case of the heavy--light system ($m_1\to\infty$, $m_2=m$).
Here $\sigma_1=\sigma_H+\eta^2(\sigma_H-\sigma_E)$, $\sigma_2=2\eta(\sigma_E-\sigma_H)$.
The fields $\mu$, $\nu(\beta)$, and $\eta(\beta)$ are the auxiliary fields, also called in the literature the einbeins
\cite{ein1}\footnote{The einbein $\mu$ is introduced in the Lagrangian via the substitution
$2\sqrt{AB}\to A/\mu+B\mu$ and allows one to simplify the quark kinetic term. The continuous einbein $\nu(\beta)$ enters through the
same trick for the string term. Finally, the second continuous einbein $\eta(\beta)$ appears due to the substitution $A^2/B\to-B\eta^2+2\eta
A$. As soon as extrema are taken in all einbeins, the initial form of the Lagrangian is restored.}.
Generally speaking, the einbein fields appear in the Lagrangian and, even in absence of the corresponding
velocities, they can be considered as extra degrees of freedom introduced to the
system. The einbeins can be touched upon when proceeding from the Lagrangian of the system to its Hamiltonian and thus they mix with the ordinary
particles coordinates and momenta. Besides, in order to
preserve the number of physical degrees of freedom, constrains are to be imposed on the system and then the formalism of constrained systems quantisation
\cite{Dirac} is operative (see, for example, Ref.~\cite{ein3} for the open straight--line QCD string quantisation using this formalism).
A nontrivial algebra of constraints and the process of disentangling the physical degrees of freedom and non-physical ones make the problem very
complicated. In the meantime, a simpler approach to einbeins exists which amounts to considering all (or some) of them as variational parameters and thus
to taking extrema in the einbeins either in the Hamiltonian or in its spectrum \cite{ein2}. Being an approximate approach this technique
appears accurate enough (see, for example, Ref.~\cite{KNS}) providing a simple but powerful and intuitive method of investigation.
In this Letter we follow the given technique, so that
extrema in all three einbeins are understood in the Hamiltonian (\ref{Hgen}).
Notice that, for $\sigma_E=\sigma_H=\sigma$, the field $\eta$ drops from the Hamiltonian and
the standard expression for the string with quarks at the ends \cite{DKS} readily comes out from Eq.~(\ref{Hgen}).

Notice that the kinetic part of the Hamiltonian (\ref{Hgen}) has a very clear structure: the radial motion of the quarks
happens with the effective mass $\mu$, whereas for the orbital motion the mass is somewhat different,
containing the contribution of the inertia of the string.

We take the extrema in $\nu(\beta)$ and $\eta(\beta)$ now, approximating $\eta(\beta)$ by a uniform in $\beta$ distribution.
Then the Hamiltonian (\ref{Hgen}) takes the form
\be
H=\frac{1}{\xi}\left(\frac{p_r^2+m^2}{\mu}+\mu\right)+V_{\rm SI}(r),
\label{Hllhl}
\ee
and the spin--independent potential reads:
\be
V_{\rm SI}(r)=\eta_0\sigma_E r+\left(\frac{1}{\eta_0}-\eta_0\right)\sigma_H r+\frac{1}{\xi} \mu y^2,\quad \eta_0=\frac{y}{\arcsin y},
\label{VSI}
\ee
with $y$ being the solution of the transcendental equation
\be
\frac{\sqrt{l(l+1)}}{\sigma_H r^2}=\frac{\xi}{4y}\left(1+\eta_0^2\left(1-\frac{\sigma_E}{\sigma_H}\right)\right)\left(\frac{1}{\eta_0}-
\sqrt{1-y^2}\right)+\frac{\mu y}{\sigma_H r}.
\label{yeqgen}
\ee
The interested reader can find the details of a similar evaluation performed for the Hamiltonian (\ref{Hgen})
with $\sigma_E=\sigma_H=\sigma$ in Ref.~\cite{regge,BS0}.

The remaining einbein $\mu$ is to be considered as the variational parameter to minimise the spectrum of the Hamiltonian (\ref{Hllhl}).
Obviously, the extremal value of $\mu$ depends on quantum numbers and acquires two contributions: one coming from the current quark mass $m$
and the other, purely dynamical, contribution coming from the mean value of the radial component of the momentum $p_r$.
It is instructive to pinpoint the difference in the potential (\ref{VSI}) below and above the $T_c$.

At small $r$'s, the potential (\ref{VSI}) turns to the centrifugal barrier
$l(l+1)/(\xi\mu r^2)$, whereas its large--$r$ behaviour differs dramatically for the temperatures
below and above the $T_c$. Indeed, the leading large-$r$ contribution to the inter-quark potential corresponds to $y\ll 1$ and,
for $T<T_c$, reads:
\be
V_{\rm conf}(r)=\sigma_E r.
\label{lin}
\ee
This is the linear confinement which is of a purely colour--electric nature and which admits angular--momentum--dependent corrections (see
Refs.~\cite{DKS,regge}).

In the deconfinement phase, at $T>T_c$, the colour--electric part of the potential (\ref{VSI})
vanishes, the leading long--range term coming from the angular--momentum--dependent part of the interaction:
\be
V_{\rm SI}(r)=\frac{3l(l+1)}{\xi^2\sigma_H r^3}+\ldots.
\label{VSI20}
\ee
Interestingly, in the deconfinement phase in absence of the confining potential,
the spin--independent interaction becomes short--ranged decreasing as $1/r^3$
at large inter-quark separations. This feature means the full compensation of the centrifugal barrier which
would naively behave as $1/r^2$ instead. The reason is obvious: at large inter-quark separations, the effective quark mass $\mu$ is to be compared to
the ``mass" of the string $\sigma r$. The bound--state problem solved in the potential (\ref{lin})
gives a large value $\langle p_r\rangle\propto \sigma_E\langle r\rangle$, so that, even for light (massless) quarks, their effective mass $\mu$
appears quite large ($\mu\gg m$).
On the contrary, for light quarks and in absence of the strong confining interaction (\ref{lin}), the values of $\mu$ are
small ($\mu\approx m$) and can be neglected as compared to the string contribution $\sigma_H r$. 
This makes the spin--dependent terms in the effective inter-quark interaction important in
this regime, as opposed to the confinement phase, where they give only small corrections to the bound states formed in the confining
potential (\ref{lin}).

In the next section we turn to the derivation of spin--dependent contributions to the inter-quark potential.

\section{Nonperturbative spin--dependent interactions}\label{SDint}

In this section we return to the Green's function of the quark--antiquark system (\ref{1}) and restore spin--dependent terms. To this end we
notice that the interaction of the quark spins with the background gluonic field is to be added at the exponent. It enters in the standard
combination $\sigma_{\mu\nu}F_{\mu\nu}$, with $\sigma_{\mu\nu}=\frac{1}{4i}(\gamma_\mu\gamma_\nu-\gamma_\nu\gamma_\mu)$, and appears under the
integral in the proper time $\tau$. After averaging over the background field, the resulting expression for the Wilson loop can be again
written in the form of Eq.~(\ref{W}),
\be
\frac{1}{N_C}\langle Tr W(C)\rangle=\exp\left(-\frac12\int_Sd\pi_{\mu\nu}(x)\int_Sd\pi_{\lambda\rho}(x')
\langle\langle Tr F_{\mu\nu}(x)\Phi(x,x')F_{\lambda\rho}(x')\Phi(x',x)\rangle\rangle\right),
\label{W2}
\ee
but with the differential $d\pi_{\mu\nu}$ containing the spinor part,
\be
d\pi_{\mu\nu}(x)=ds_{\mu\nu}(x)-i\sigma_{\mu\nu}d\tau.
\ee

The nonperturbative spin--dependent interaction appears from the combination of differentials involving $\sigma_{\mu\nu}$.
We skip the details of the derivation which can be found, for example, in Refs.~\cite{242,Lisbon} and quote the result here.
The leading
spin--dependent term is the spin--orbit interaction (we omit contributions of $D_1^{E,H}$ which bring about
short--range terms ${\cal O}(r^{-3})$),
\be
V_{SO}(r)=\left(\frac{\vec{S}_1\vec L_1}{2\mu_1^2}-\frac{\vec{S}_2\vec L_2}{2\mu_2^2}\right)
\left(\frac{1}{r}\frac{d V_0}{dr}+\frac{2}{r}\frac{dV_1}{dr}\right),
\label{Vls1}
\ee
where $V_i(r)$ can be expressed through the colour--electric and colour--magnetic field correlators,
\be
\frac{1}{r}\frac{d V_0}{dr}=\frac{2}{r}\int_0^\infty d\tau\int_0^r d\lambda D^E(\tau,\lambda),\quad
\frac{2}{r}\frac{dV_1}{dr}=-\frac{4}{r}\int_0^\infty d\tau\int_0^r d\lambda\left(1-\frac{\lambda}{r}\right)D^H(\tau,\lambda).
\label{SIall}
\ee

Notice that this result \cite{242,Lisbon}, is not due to the $1/m$ expansion, but is obtained with the only
approximation made being the Gaussian approximation for field correlators. Accuracy of this approximation was checked both at $T=0$ \cite{30}
and at $T>T_c$ \cite{24} to be of the order of one percent.

Only the $V_1$ potential survives above the $T_c$. Besides
the corresponding $\mu$--dependent denominator is to be corrected according to the discussion of the
previous section --- namely, one of $\mu$'s is to be augmented by the string rotation term yielding
\be
V_{SO}(r)=-\frac{2\xi\vec{S}\vec{L}}{\mu r(\xi\mu+2\langle\nu(\beta-\xi/2)^2\rangle)}\int_0^\infty d\tau\int_0^r
d\lambda D^H(\tau,\lambda)\left(1-\frac{\lambda}{r}\right),
\label{SOgen}
\ee
where
\be
\langle\nu(\beta-\xi/2)^2\rangle\equiv\int_0^1d\beta\nu(\beta-\xi/2)^2=
\frac{\sigma_H r}{2\xi^2y^2}\left(1+\eta_0^2\right)\left(\frac{1}{\eta_0}-\sqrt{1-y^2}\right),
\label{yeqgen2}
\ee
and $\vec{S}=\vec{S}_1+\vec{S_2}$, for the light--light system, and $\vec{S}$ is the light--quark spin, for the heavy--light quarkonium.

\section{Bound states of heavy quarks above the deconfinement temperature}

\begin{figure}[t]
\centerline{\epsfig{file=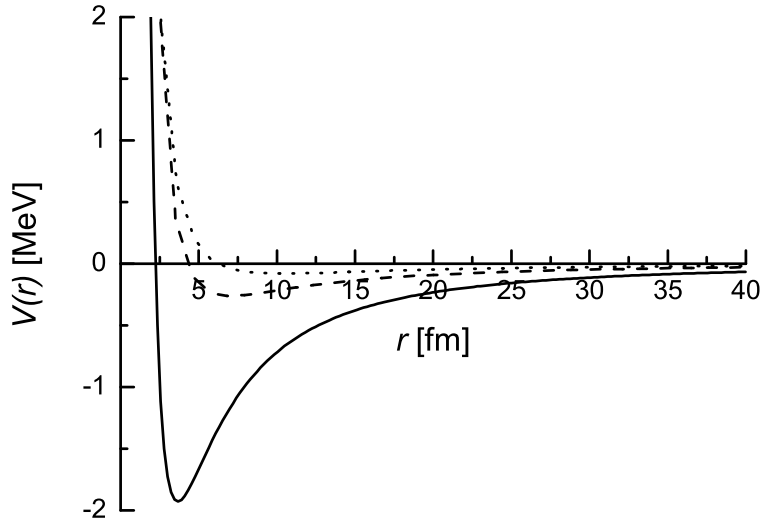,width=10cm}}
\caption{The profile of the effective potential (\ref{Vf}) for $m=1$GeV (solid line), $m=2$GeV (dashed line), and
$m=3$GeV (dotted line).}\label{pots}
\end{figure}

In the previous sections we derived the effective nonperturbative
inter-quark potential, including spin--in\-de\-pen\-dent terms and the spin--orbital interaction.
It follows from Eq.~(\ref{SOgen}) that above the deconfinement temperature,
for the states with the total momentum $J=l+S$, $\vec{S}\vec{L}>0$ and the potential $V_{SO}(r)$
becomes attractive, with a possibility to maintain bound states. Furthermore, its
slow decrease as $r\to\infty$ suggests that an infinite number of bound states exists, with the binding energies asymptotically
approaching zero. Let us study these bound states in more detail. Hereafter $\sigma_E=0$ and we use the notation
$\sigma$ for the magnetic tension $\sigma_H$.

In view of an obvious similarity of the light--light and heavy--light cases
(the difference manifesting itself only in numerical coefficients), we investigate numerically
only the light--light system, as a paradigmatic example. Furthermore, for $r\gg T_g$, the potential (\ref{SOgen}) does not depend on
the form of the correlator $D^H$ since
\be
2\int_0^\infty d\tau\int_0^r d\lambda D^H(\tau,\lambda)\left(1-\frac{\lambda}{r}\right)
\mathop{\approx}\limits_{r\gg T_g}\sigma.
\ee
Finally, we neglect the perturbative part of the inter-quark interaction for it is screened to a large extend contributing to short--ranged
forces only whereas the effect discussed in this work is essentially a long--ranged effect.

Therefore we study the spectrum of bound states in the potential
\be
V(r)=\left(\frac{\arcsin y}{y}-\frac{y}{\arcsin y}\right)\sigma r+\mu y^2
-\frac{\sigma l}{\mu r(\mu+2\langle\nu(\beta-1/2)^2\rangle)},
\label{Vf}
\ee
which is the sum of the spin--independent term (\ref{VSI}) and the spin--orbital term (\ref{SOgen});
$y$ is the solution of Eq.~(\ref{yeqgen}) with $\sigma_E=0$.
In Fig.~\ref{pots} we plot the effective potential (\ref{Vf}) for three values of the quark mass:
$m=1$GeV, $2$GeV, and $3$GeV.

The resulting eigenenergy $\varepsilon_{n_rl}(\mu)$ is added then to the free part of the
Hamiltonian (\ref{Hgen}),
\be
M_{n_rl}(\mu)=\frac{m^2}{\mu}+\mu+\varepsilon_{n_rl}(\mu),
\label{Mmin}
\ee
and this sum is minimised with respect to the einbein $\mu$,
\be
\left.\frac{\partial M_{n_rl}(\mu)}{\partial\mu}\right|_{\mu=\mu_0}=0,\quad M_{n_rl}=M_{n_rl}(\mu_0).
\ee

In Table~1 we present the set of parameters used in our numerical calculations,
whereas in Table~2 we give the results for the binding energy 
for the $b\bar{b}$, $c\bar{c}$, and $s\bar{s}$ quarkonia above the $T_c$ for $l=1$ and $n_r=0,1$.
We ensure therefore that for $l\neq 0$ the potential (\ref{Vf}) does support bound states. The binding
energy is small ($|E_{n_rl}|\ll T$ for the $b$ and $c$ quarks and $|E_{n_rl}|\lesssim T$ for $s$ quarks) so these bound
states can dissociate easily.

\begin{table}[t]
\begin{center}
\begin{tabular}{|c|c|c|c|c|c|}
\hline
Parameter&$m_b$, GeV& $m_c$, GeV& $m_s$, GeV&$\sigma$, GeV$^2$& $T_g$, fm\\
\hline
Value&4.8&1.44&0.22&0.2&0.2\\
\hline
\end{tabular}
\end{center}
\caption{The set of parameters used for the numerical evaluation.}
\end{table}

\begin{table}[t]
\begin{center}
\begin{tabular}{|c|c|c|c|}
\hline
&$b\bar{b}$&$c\bar{c}$&$s\bar{s}$\\
\hline
$n_r=0$&-0.007&-0.19&-45\\
\hline
$n_r=1$&-5$\times 10^{-4}$&-0.015&-2.7\\
\hline
\end{tabular}
\end{center}
\caption{The binding energy $E_{n_rl}\equiv M_{n_rl}-2m$ (in MeV) for the ground state and for the first radial excitation
in the potential (\ref{Vf}) with $l=1$ for the $b\bar{b}$, $c\bar{c}$, and $s\bar{s}$ quarkonia.}
\end{table}

\section{Bound states of light quarks above the deconfinement temperature}

\begin{figure}[t]
\begin{tabular}{cc}
\epsfig{file=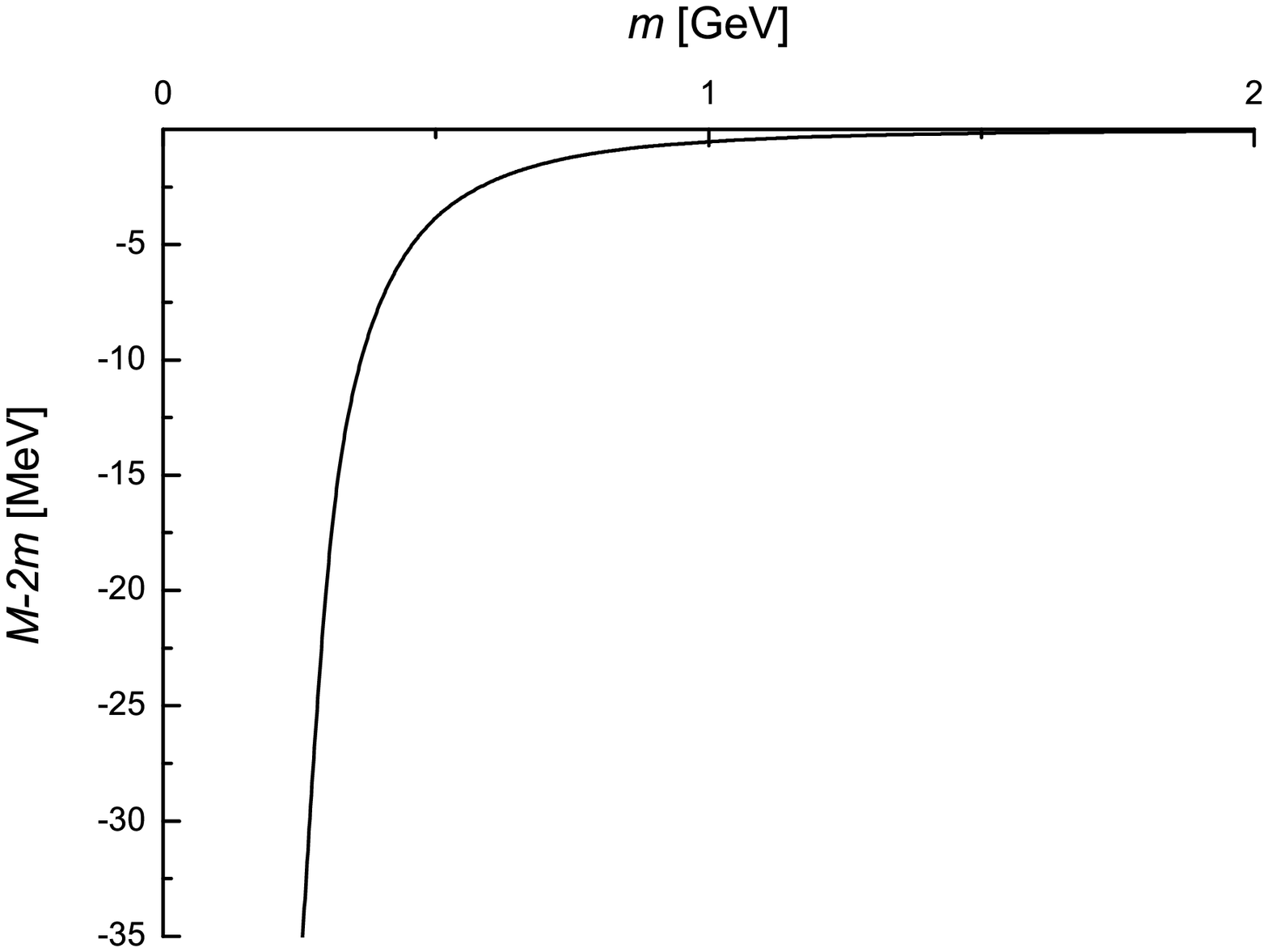,width=7.9cm}&\epsfig{file=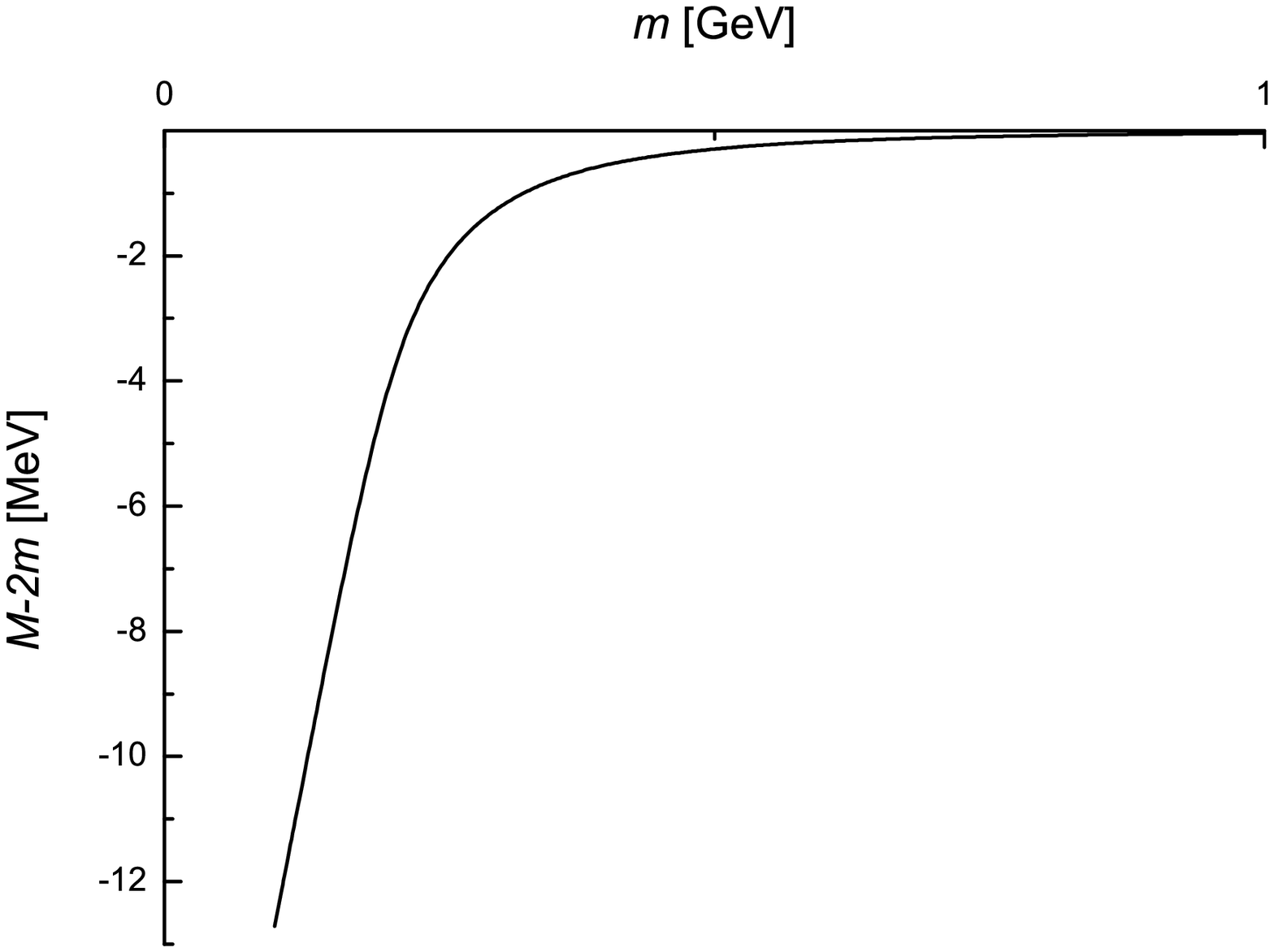,width=7.9cm}
\end{tabular}
\caption{The binding energy of the quark--antiquark system versus the mass of the quark
for $l=1$ and $n_r=0$ (first plot) and $n_r=1$ (second plot).}\label{masses}
\end{figure}

In this section we turn to the problem of binding of light quarks.
The effective potential (\ref{Vf}) admits different forms at different inter-quark separations,
depending on which contribution, of the quark mass term  $\mu$ or of the ``string mass"
$2\langle\nu(\beta-1/2)^2\rangle)$,
gives the dominating contribution, that is for $\mu\gg\sigma r$ and $\mu\ll\sigma r$.
If large distances contribute most to the bound state
formation (the latter case), then $V(r)={\cal O}(l(l+1)/(\sigma r^3))+{\cal O}(l/(\mu r^2))$, where the first
term comes from the spin--independent interaction (see Eq.~(\ref{SIall})) and the other stems from the
spin--orbit potential. The dependence of the binding energy on $\mu$ is expected then to be rather
moderate, approximately as $1/\mu$.

On the contrary, in the former case with the string dynamics giving a correction to the quark mass term,
the potential (\ref{Vf}) can be approximated as
\be
V(r)\approx\frac{l(l+1)}{\mu r^2}-\frac{\sigma l}{\mu^2 r},
\ee
that is by the sum of the centrifugal barrier and the attractive Coulomb-like potential with the effective coupling
\be
\alpha_{\rm eff}=\frac{\sigma l}{\mu^2}.
\label{aeff}
\ee
The corresponding eigenenergy can be found in any textbook in Quantum Mechanics and gives a stronger dependence on $\mu$,
\be
\varepsilon(\mu)\propto -\mu\alpha_{\rm eff}^2\propto -\frac{\sigma^2 l^2}{\mu^3}.
\label{mucoul}
\ee

Let us consider the states with $l=1$ and $n_r=0$.
We follow now the procedure described in detail in the previous section, solve the full problem numerically,
and find the dependence of the eigenenergy $\varepsilon$ on the einbein $\mu$ to be
\be
\varepsilon(\mu)\propto -\frac{1}{\mu^{2.79}}.
\label{vemu}
\ee

Comparing this to Eq.~(\ref{mucoul}) we find a good agreement, with the small deviation in the power resulting
from the proper string dynamics. We conclude therefore that the dynamics of the system develops at the
inter-quark separations $T_g\ll r\lesssim m/\sigma$ (since, for the set of parameters given in Table~1, $m\gtrsim\sigma T_g$ then there is room for such separations).

Let us discuss now the procedure of minimisation of the spectrum (\ref{Mmin}) in $\mu$. First of all,
let us notice that, in the einbein field formalism, the calculation of the
spectrum naively looks like a nonrelativistic calculation due to the ``nonrelativistic" form of the kinetic energy in
the Hamiltonian with the einbein field $\mu$ introduced. In the meantime, the full relativistic form of the
quark kinetic energy is readily restored as $\mu$ takes its extremal value and hence this is the procedure of
taking extremum in $\mu$ in the masses (\ref{Mmin}) to sum up an infinite series of relativistic corrections and thus 
to restore the relativistic spectrum. For example, the relativistic
ground state eigenenergy $E_0=m\sqrt{1-(Z\alpha)^2}$ of the one--body Dirac equation with the Coulomb potential $-Z\alpha/r$ can be reproduced {\em exactly} with the help of the einbein technique.
Finally, one can visualise the form of $\mu$ considering the effective
Dirac equation for the light quark in the field of the static antiquark source. When written in the form of a second--order
differential equation, it
contains the spin--orbit term of the same form as given in Eq.~(\ref{Vls1}) but with $\mu$ replaced by the
combination $\epsilon+m+U-V$, where $U$ and $V$ are scalar and vector potentials,
respectively. For light (massless) quarks this combination takes drastically different values below and above the deconfinement temperature. Indeed, in the confining phase of QCD, when spontaneous chiral symmetry breaking leads to a strong effective, dynamically generated scalar potential $U$, this effective ``$\mu$" is large. On the contrary, above the $T_c$, when $U$ is small ``$\mu$" is also small (it can be even negative since the eigenenergy $\epsilon$ may have any sign).

We find numerically that the extremum in $\mu$ for $M_{n_rl}(\mu)$, Eq.~(\ref{Mmin}),
exists for $m$ exceeding the value of approximately $0.22$GeV (for the given $\sigma=0.2$GeV), and no extremum
exists for smaller values of the quark mass (see Fig.~\ref{masses} for the dependence of the binding energy
$E_{n_rl}$ on the quark mass). This property of the bound state spectrum can be easily
understood using the analogy with the bound state problem for the Dirac equation with the potential in the form of a
deep square well or Coulomb potential discussed above.
For example, for the Coulomb potential, a problem appears as the coupling
exceeds unity --- the well--known problem of $Z>137$. From Eq.~(\ref{aeff}) we easily find this critical
phenomenon to happen at $m\approx\mu\lesssim \sqrt{\sigma}\approx 0.4$GeV.
This estimate is in good agreement with the result of
our direct numerical calculations quoted above.

Physically this situation means that many quark--antiquark and/or gluon pairs are formed and finally stabilise
the vacuum. Formally the problem is not anymore a two--body problem, but rather many--body, so that
many--body techniques are to be applied. For example, in electrodynamics with $Z>137$,
one can derive the resulting self-consistent field of the Thomas--Fermi type \cite{z137}.
A similar situation can be expected in the deconfinement phase of QCD. In absence of the linear potential,
the einbein $\mu$ (playing the role of the effective quark mass) is not anymore bounded from below by the
values of order $\sqrt{\sigma_E}\simeq 0.4$GeV coming from the binding energy in the linearly rising potential.
To see the onset of this phenomenon in the framework of our two--body (one--body for the heavy--light case)
Hamiltonian, one should take into account the negative--energy part of the spectrum, when the full matrix
form of the Hamiltonian is considered \cite{negmu}. Indeed, the matrix structure of the Hamiltonian occurs
in the path--integral formalism from the two--fold time--forward/backward motion described by the
positive/negative values of $\mu$. Off--diagonal terms in the matrix Hamiltonian produce the turning
points in the particle trajectory and result in Z--graphs.

Notice that the same is true for the glueballs and gluelumps since in this case
equations are the same as for light--light and heavy--light quarkonia, respectively, but with the quark spin
replaced by the gluon spin and $\sigma_H$ by $\frac94\sigma_H$.

Concluding this section one can say that colour--magnetic (spin--dependent) interaction acting on light quark
or gluonic systems enforces nonperturbative creation of light $q\bar{q}$ and $gg$ pairs.

\section{Discussion}

The results obtained in this Letter allow us to comment on the general situation with the existence of bound quark--antiquark states in the
deconfinement phase of QCD. First of all, contrary to naive expectations, the colour--Coulomb potential is screened down to a
short--ranged interaction and  bound states appear due
to nonperturbative colour--electric (see Refs.~\cite{8,9}) and
colour--magnetic interactions in the vacuum. Indeed, although quark--antiquark pairs in the
relative S--wave cannot be bound by such
interactions for $T\gtrsim 1.5T_c$ (see, for example, Refs.~\cite{8,9,10}), pairs with a nonzero relative angular momentum can form bound states at
all $T>T_c$ since $\sigma_H$ grows with $T$.
Second, formation of such bound states above the $T_c$ is energetically favourable since it lowers the system energy as compared to the
ensemble of free, unbound particles. For heavy quarks the binding is weak and the system dissociates easily.
Finally, the dependence of the binding energy on the quark mass is strong --- the corresponding eigenenergy
for strange quarks is around $10\div 100$MeV rather then below 1MeV for the charmed and bottom quarks
(see Table~2).
The situation becomes even more dramatic for the lightest quarks.
The effective inter-quark potential for light quark flavours (and gluons!) becomes extremely strong and may lead to 
pair creation --- the effect similar to the critical phenomenon in QED for the centre charge $Z$ exceeding 137.
Vacuum polarisation effects become important and they lead to a complete rebuilding of the vacuum structure of the theory.
We anticipate a similar phenomenon to take place for light quarks (and gluons) in QCD in the deconfinement phase.

It is important to notice that it was a separated quark--antiquark pair which was considered in this paper. In reality such quark--antiquark
pairs are to be considered in the medium formed by other quarks and gluons, that is as a part of the SQGP. As a measure of the
interaction in SQGP one can consider the ratio of the mean potential energy to the mean kinetic energy of the particles in the plasma,
$\Gamma=\langle V\rangle/\langle K\rangle$. It is easy to estimate that $\langle K\rangle\simeq T$ and
$\langle V\rangle\simeq \sigma_H/T$.
This gives $\Gamma=\sigma_H/T^2$ and so this parameter is large for quarks and it is several times
larger for gluons. Therefore, SQGP is a strongly interacting
medium which looks like a liquid, rather then as a gas.
With the growth of the temperature the
medium becomes more dense, and the mean distance between particles decreases.
As this distance becomes comparable to the radius of the bound
states discussed in this Letter, the latter will dissociate because of the screening effects.
In other words, the hot medium plays the role
of a natural cut--off for the effect of bound pair creation discussed above.
Notice however that, for the quark masses around 0.2GeV, the
radius of the bound state is of the order of  one fm and it is expected to
decrease further with the decrease of the quark mass, even if the pair
creation process is properly taken into account. This means that indeed there is room for such bound states for the temperatures above the
$T_c$. Breakup, with the growth of the temperature, of such high--$l$
states for
quarks, and especially for gluons which possess  more degrees of freedom than quarks, may affect such
characteristics of the plasma as its free energy and it entropy
(for a recent attempt of explaining the near--$T_c$ behaviour of these characteristics see Ref.~\cite{ant}).
This work is in progress and will be reported elsewhere.
\medskip

The authors would like to acknowledge enlightening discussions with K. G. Boreskov, A. B. Kaidalov, and O. V. Kancheli.
This research was supported by the Federal Agency for Atomic Energy of Russian Federation, by the grants RFFI-06-02-17012,
RFFI-05-02-04012-NNIOa, DFG-436 RUS 113/820/0-1(R), NSh-843.2006.2, by the Federal Programme of
the Russian Ministry of Industry, Science, and Technology No.
40.052.1.1.1112. A.N. is also supported through the project PTDC/FIS/70843/2006-Fisica.

\end{document}